\documentclass{iau}
\usepackage{graphicx}

\title[Spectroscopy of fullerenes, fulleranes and PAHs] 
{Spectroscopy of fullerenes, fulleranes and PAHs in the UV, visible and
near infrared spectral range}

\author[Cataldo et al.]   
{Franco Cataldo$^{1,2}$, D. A. Garc\'{\i}a-Hern\'andez$^{3,4}$, Arturo
Manchado$^{3,4,5}$ and Susana Iglesias-Groth$^{3,4}$
\thanks{This research work has been supported by grant AYA$-$2007$-$64748 Expte.
NG$-$014$-$10 of the Spanish Ministry of Science and Innovation (MICINN). We
also acknowledge support provided by the Spanish Ministry of Economy and
Competitiveness under grant AYA$-$2011$-$27754.}
}

\affiliation{$^1$Istituto Nazionale di Astrofisica - Osservatorio Astrofisica di Catania, \\ Via S. Sofia 78, 95123, Italy; email: {\tt franco.cataldo@fastwebnet.it} \\[\affilskip]
$^2$Actinium Chemical Research, Via Casilina 1626/A, 00133 Rome, Italy\\[\affilskip]
$^3$Instituto de Astrof\'{\i}sica de Canarias, C/ Via L\'actea s/n, E$-$38200 La Laguna, Tenerife, Spain \\[\affilskip]
$^4$Departamento de Astrof\'{\i}sica, ULL, E$-$38206 La Laguna, Tenerife, Spain\\[\affilskip]
$^5$CSIC, Madrid, Spain
}

\pubyear{2013}
\volume{297}  
\pagerange{119--126}
\jname{The Diffuse Insterstellar Bands}
\editors{Jan Cami \& Nick Cox, eds.}
\begin{document}

\maketitle

\begin{abstract}
The spectra of fullerenes C$_{60}$ and C$_{70}$, higher fullerenes C$_{76}$,
C$_{78}$ and C$_{84}$ and hydrogenated fullerenes (fulleranes) were studied in
laboratory in the UV and in the visible spectral range and could be used for
searching and recognizing these molecules in space. Furthermore, the radical
cation spectra of all the mentioned fullerene series and also of a series of
large and very large polycyclic aromatic hydrocarbons (PAHs) were generated in
laboratory and studied in the near infrared spectral range.
\end{abstract}


C$_{60}$ and C$_{70}$ fullerenes display quite complex spectra in the
ultraviolet and visible spectral regions due to the electronic transitions of
the $\pi$ electrons of their weakly conjugated double bonds. C$_{60}$ shows two
intense bands at 213 and 257 nm, respectively, with a molar extiction
coefficient $\varepsilon$ of 135000  and 175000 L cm$^{-1}$ mol$^{-1}$,
respectively. Other bands of C$_{60}$ appear at 329 nm ($\varepsilon$ = 51000 L
cm$^{-1}$ mol$^{-1}$), 404 nm and a broad and relatively weak band between 440
and 670 nm with subfeatures at 500, 540, 570, 600 and 625 nm. Similarly, also
the electronic absorption spectrum of C$_{70}$ is characterized by two intense
transitions at 210 and 228 nm followed by a series of weak electronic
transitions at 327, 358 and 375 nm and by a weak and broad band at 460 nm. Both
C$_{60}$ and C$_{70}$ fullerenes are extremely reactive and avid of atomic
hydrogen and it was demonstrated experimentally that C$_{60}$ is transformed
very quickly into the fullerane C$_{60}$H$_{36}$ when reacted with atomic
hydrogen while C$_{70}$ yields C$_{70}$H$_{38}$  (Cataldo \& Iglesias-Groth
2009, 2010). The same reactivity is also shown by C$_{60}$ and C$_{70}$ with
deuterium yielding the deuterated fulleranes analogous to the hydrogenated
derivatives mentioned above (Cataldo et al. 2009a,b). The C$_{60}$H$_{36}$
fullerane displays an electronic absorpion spectrum with a unique absorption
band at 217 nm with a molar extinction coefficient $\varepsilon$ = 17140 L
cm$^{-1}$ mol$^{-1}$ while C$_{60}$D$_{36}$ shows an analogous spectrum with a
maximum at 217 nm and $\varepsilon$ = 16480 L cm$^{-1}$ mol$^{-1}$ (Cataldo \&
Iglesias-Groth 2009; Cataldo et al. 2009c). It has been shown that the peak of
C$_{60}$H$_{36}$ matches both in position and in width the UV bump of the
interstellar light extinction curve (Cataldo \& Iglesias-Groth 2009). It is
interesting to note that also the hydrogenated and deuterated C$_{70}$ under the
form of C$_{70}$H$_{38}$ and C$_{70}$D$_{38}$ display an analogous UV spectrum
as that shown by the hydrogenated and perdeuterated C$_{60}$ derivatives. The
peak position is found at 214 nm with $\varepsilon$ = 6300 L cm$^{-1}$
mol$^{-1}$ in the case of C$_{70}$H$_{38}$ and $\varepsilon$ = 5800 L cm$^{-1}$
mol$^{-1}$ in the case of C$_{70}$D$_{38}$ (Cataldo et al. 2009d). These results
imply that hydrogenated and deuterated C$_{60}$ and C$_{70}$ fullerenes can be
present in the interstellar medium and may contribute to the UV bump of the
interstellar light extinction curve and may play a key role in molecular
hydrogen formation in space starting from the atomic hydrogen adsorption on the
fullerene cage structure and release as molecular hydrogen under UV irradiation
(Cataldo \& Iglesias-Groth 2010). Indeed, it was shown experimentally (Cataldo
\& Iglesias-Groth 2009; Cataldo et al. 2009c,d) that the UV irradiation of
fulleranes, the hydrogenated fullerenes, causes the release of molecular
hydrogen and the consequent partial dehydrogenation of the fullerane with a
shift of the electronic transition at 217 nm toward longer wavelengths. Another
characteristic property of hydrogenated/deuterated fullerenes regards the
kinetic isotope effect observed experimentally during the photolysis, which
involves an easier photolysis of the C-H bond in comparison to the C-D bond with
the consequent possible H/D fractionation on the fullerene surface (Cataldo \&
Iglesias-Groth 2009; Cataldo et al. 2009c,d).

In the interstellar medium under the action of cosmic rays or other radiation,
C$_{60}$ may undergo ionization losing one electron and yielding a radical
cation. In laboratory the radical cation spectrum of C$_{60}$ was produced in a
very high oxidizing medium known as oleum (Cataldo et al. 2012a) and the
resulting electronic transitions were found in the near-infrared at 823.7 nm and
at 931.5 nm. The latter transition was interpreted as the true radical cation
transition of C$_{60}$. On the other hand, under the same conditions the radical
cation of C$_{70}$ displays its longest wavelength transition at about 640 nm
(Cataldo et al. 2012a). When hydrogenated C$_{60}$ and C$_{70}$ are dissolved in
the same medium that generates C$_{60}$$^{+.}$ and C$_{70}$$^{+.}$, it happens
that they undergo an almost complete dehydrogenation, displaying once again the
bands of the radical cation spectra (Cataldo et al. 2009b, 2012b):
C$_{60}$H$_{36}$ $+$ oleum $\rightarrow$ C$_{60}$$^{+.}$ $+$ 18H$_{2}$ and
C$_{70}$H$_{38}$ $+$ oleum $\rightarrow$ C$_{70}$$^{+.}$ $+$ 19H$_{2}$.

An attempt to record the radical cation spectra of higher fullerenes like
C$_{76}$, C$_{78}$ and C$_{84}$ in oleum was also made but the results show only
weak and broad transitions in the near infrared that may suggest that higher
fullerenes once oxidized to the radical cation stage undergo further oxidative
decomposition (at least in oleum) (Cataldo et al. 2012a).

Although less probable than the radical cation formation, it is expected that in
space C$_{60}$ and C$_{70}$ fullerenes may produce also the radical anion
(Cataldo et al. 2013a). The radical anion spectrum of C$_{60}$ is characterized
by a complex band pattern, showing a series of peaks at 935, 994, 1032 and 1058
nm while the radical anion spectrum of C$_{70}$ shows only a broad band with a
peak in the range comprised between 814 and 840 nm.

The infrared spectra of a series of reference fullerenes and fulleranes have
been recorded in the laboratory by Iglesias-Groth et al. (2011, 2012) together
with the relative molar absorptivity. This recent laboratory spectroscopy of 
C$_{60}$ and C$_{70}$ fullerenes has permitted an accurate determination of the
fullerene abundances in space (Garc\'{\i}a-Hern\'andez et al. 2011, 2012). In
addition, because the reference spectra of hydrogenated fullerenes (fulleranes)
are now available, a search for these molecular species can now be made in
space, and their possible detection in astrophysical environments may be only a
matter of time and luck.

Very large polycyclic aromatic hydrocarbons (VLPAHs) are not easily accessible
but of high interest as reference molecules for the explanation of the diffuse
interstellar bands (DIBs). Using the Scholl reaction, Cataldo et al. (2011) have
synthesized a series of VLPAHs, ranging from dicoronylene to quaterrylene to
hexabenzocoronene. Dicoronylene was also obtained by the thermal dimerization of
coronene. If the thermal treatment of coronene is prolonged, then the
oligomerization of coronene proceeds further, yielding a black oligomer, which
is probably a trimer or a higher homologue. It is shown that from coronene
oligomerization it is possible to build a sheet of graphene. 

The radical cation spectra of a series of normal PAHs and VLPAHs were studied by
Cataldo et al. (2010) using oleum as oxidizing medium. PAHs and VLPAHs in the
harsh space conditions are expected to undergo ionization, being mainly in the
form of radical cations. The radical cation spectra of PAHs and VLPAHs are
characterized by intense electronic transitions, which in our laboratory
conditions appear as relatively broad bands but in the gas phase and at
relatively lower temperature such electronic transition must appear as sharp and
narrow bands. Thus, it is possible that some DIBs could be associated to the
radical cation transitions of some PAHs or, even better, to some VLPAHs. It
could be useful to mention here some of the results reported in the work of
Cataldo et al. (2010). First of all the PAHs and VLPAHs once oxidized to the
radical cation display electronic transitions in the visible and in the near
infrared and may be associated to certain DIBs. The anthracene radical cation
shows its longest wavelength transition (LWT) at 588.2 nm, tetracene radical
cation LWT = 652.8 nm, pentacene radical cation LWT = 760.4 nm, pyrene radical
cation LWT = 556 and 602 nm, coronene radical cation LWT = 574 nm, dicoronylene
radical cation LWT = 755 nm, coronene oligomer radical cation LWT = 881 nm,
perylene radical cation LWT = 515 nm and quaterrylene radical cation LWT = 811
and 838 nm. Finally, far infrared spectroscopy on a series of PAHs, VLPAHs and
petroleum and coal fractions were also reported in recent works by Cataldo et
al. (2013b,c).

\end{document}